\documentstyle[emulateapj]{article}

\newcommand\etal{et\thinspace al.~}

\newcommand\msol{\rm\,M_\odot}
\newcommand\zsol{\rm\,Z_\odot}

\newcommand\nins{N_{\rm ins}(z)}
\newcommand\ntot{N_{\rm tot}(z)}
\def\spose#1{\hbox to 0pt{#1\hss}}
\def\simpropto{\mathrel{\spose{\raise 3pt\hbox{$\propto$}}
     \lower 3.0pt\hbox{$\sim$}}}

\received{7 July 2000}
\accepted{18 August 2000}

\slugcomment{Accepted by ApJ Letters, 18 August 2000}

\lefthead{M. S. Oey}
\righthead{Inhomogeneous chemical evolution}

\begin{document}

\title{	
A new look at simple inhomogeneous chemical evolution
}

\author{M. S. Oey\altaffilmark{1}}
\altaffiltext{1}{STScI Institute Fellow.}
\affil{Space Telescope Science Institute, 3700 San Martin Drive,
	Baltimore, MD   21218, USA; oey@stsci.edu}

\begin{abstract}
A rudimentary, one-zone, closed-box model for inhomogeneous chemical
evolution is offered as an alternative reference than the Simple model
in the limit of no mixing.  The metallicity distribution functions
(MDFs) of Galactic halo and bulge stars can be matched by 
varying a {\it single} 
evolutionary parameter, $nQ$.  $Q$ is the filling factor of
contaminating regions and $n$ is the number of star-forming generations. 
Therefore, {\it Q and n have equivalent roles}, and
combinations of $n$ and $Q$ yield systems with different
metallicities at any given age.  The model also revises interpretation
of observed MDFs.  Unevolved systems probe 
the parent distribution of metal production $f(z)$, for example, the
high-metallicity tail of the halo distribution agrees with a power-law
$f(z)$.  
\end{abstract}


\section{Introduction}

The relation between galactic star formation history and the
interstellar medium (ISM) is recorded in the distribution of elements
we observe today.  The principal signatures are:  (A) Time-integrated
metallicity distribution functions (MDFs), as exhibited by long-lived
stars; (B) instantaneous MDFs at a given time, for example, the
present-day dispersion in the inhomogeneous ISM; (C) abundance
ratios of elements with varying origin and histories; and
(D) age-metallicity relations.  A unified
understanding of these tracers is fundamental
to solving the puzzle of galaxy assembly and evolution.

Over the 30 years of work on chemical evolution modeling, the constraints
offered by (B), the instantaneous MDF, have been relatively neglected.
However, Tinsley (1975) showed early on that chemical inhomogeneity 
significantly affects the other chemical signatures.
Edmunds (1975) also argued that simplistic inhomogeneous
evolution of overlapping supernova remnants, with no mixing, yields a
present-day ISM that is orders of magnitude {\it smoother} than what is
presently observed.  This issue is only rarely discussed (e.g.,
Roy \& Kunth 1995).  But recently, the significance of dispersions in
the chemical signatures has been reemphasized (e.g., Wyse 1995; van
den Hoek \& de Jong 1997).  We are therefore overdue to reexamine
inhomogeneous chemical evolution. 

Picking up threads from the early work, I present here
an analytic model that predicts the metallicity dispersion from first
principles in the limiting case of no mixing.
As a reference, we return to the standard, Simple model (e.g., Tinsley 1980;
Pagel \& Patchett 1975; Schmidt 1963), which assumes: 
(1) a closed system, (2) initially metal-free,
(3) constant stellar initial mass function (IMF), and
(4) chemical homogeneity at all times.
Most previous analytic investigations of inhomogeneous 
evolution were tied to the Simple model, and assumed a given
dispersion in metal production: 
Tinsley (1975) adopted a fixed metallicity dispersion and propagated
this through the Simple model; while Searle (1977) and Malinie et
al. (1993) considered an ensemble of regions that individually 
follow the Simple model such that they yield a given dispersion.

\section{A fresh approach}

A different, perhaps simpler, approach, returns to the concept of
overlapping regions of 
contamination, as discussed qualitatively by, e.g, Edmunds (1975).
We adopt the tenets, except for (4), of the Simple
model.  Consider an initially metal-free ISM, in which a first
generation of star forming regions is randomly distributed, occupying
a volume filling factor $Q$.  We assume that these individual 
regions have a distribution of metal production $f(z)$, which is
the probability density function for obtaining metallicity $z$ at any
point affected by star formation.
Note that it is not necessary to assume anything specific about the
sizes of the individual regions, but only the relative total volumes of
contaminated to primordial gas, given by $Q$.

We now consider $n$ subsequent generations of star formation.  The individual
regions also fall randomly in the ISM, but for each generation, occupy
the same filling factor $Q$ as the first.  The role of this last assumption
will become apparent in \S 4 below.
The probability that any given point in the ISM is
occupied by $j$ overlapping regions is given by the binomial
distribution in $n$ and $Q$:
\begin{equation}\label{binomial}
P_j = \Biggl({n \atop j}\Biggr)\ Q^j\ (1-Q)^{n-j} \ \ ,\quad
	1 \leq j \leq n \quad ;
\end{equation}
and the probability of a point retaining primordial metallicity is,
\begin{equation}
P_0 = (1 - Q)^n \quad .
\end{equation}
For our purposes, we define the metallicity $z$ to be the mass
of metals per unit volume; the coexisting mass density of
H is assumed to be spatially uniform.
In accordance with our assumption of no mixing, we assume that the
metallicity at any given point is the sum of those contributed by the
individual overlapping regions $z_i$:
\begin{equation}\label{zsum}
z = \sum_{i=1}^{j}\ z_i \quad .
\end{equation}

To obtain the instantaneous MDF after $n$ generations,
we sum the distributions $N_j(z)$ for individual
sets of regions occupied by $j$ objects, weighted by their $P_j$:
\begin{equation}\label{nz_gen}
\nins = \sum_{j=1}^{n}\ P_j\ N_j(z) \quad .
\end{equation}

To derive the total MDF of all objects ever created
over $n$ generations, we must reduce the total gas mass
after each generation $k$ by a factor $D_k = 1-k\delta$, to account
for the conversion of gas into stars.  We assume the reduction
increment $\delta$ to be constant, so that the present-day gas
fraction $\mu_1 = 1-n\delta$.  The total, time-integrated MDF
is then:
\begin{equation}\label{ntot_gen}
\ntot =  \frac{1}{n}\ \sum_{k=1}^{n}\ D_{k-1}\ N_{\rm ins,k}(z)
	= \frac{1}{n}\ \sum_{j=1}^{n}\ \sum_{k=j}^{n}\ D_{k-1}\ P_{j,k}\ N_j(z) 
	\ ,
\end{equation}
where $P_{j,k}$ is $P_j$ for a given generation $k$.  Note that this
treatment for gas consumption assumes that $P_{j,k}$ remain unaffected
by the reduction of gas.

We now assume that $f(z)$ remains the same for all generations.
For large $j$, the Central Limit Theorem applies, which
predicts that the summed MDF approximates a normal distribution with mean
and variance given by $j$ times the mean $a$ and variance $\sigma^2$ 
of $f(z)$.  Thus, $\nins$ is given simply by:
\begin{equation}\label{nz}
\nins = \sum_{j=1}^{n}\ P_j\ (2\pi j\sigma^2)^{-1/2} \ 
	e^\frac{-(z-ja)^2}{2j\sigma^2}  \quad ,
\end{equation}
where $P_j$ is given by equation~\ref{binomial}.
Likewise, to estimate $\ntot$ for large $k$, the binomial distribution
approximates a normal distribution with mean and variance given by
\begin{eqnarray} \label{binomstats}
a_{\rm b} & = & kQ \nonumber \\
\sigma_{\rm b}^2 & = & kQ(1-Q) \quad .
\end{eqnarray}
Thus equation~\ref{ntot_gen} may be written:
\begin{equation}\label{ntot}
\ntot = \frac{1}{n}\ \sum_{j=1}^{n}\ \sum_{k=j}^{n}
	\frac{D_{k-1}}{j^{1/2}\ 2\pi \sigma_{\rm b}\sigma}\ 
	e^{\frac{-(j-a_{\rm b})^2}{2\sigma_{\rm b}^2}\ +
	\frac{-(z-ja)^2}{2j\sigma^2}}  \ \ .
\end{equation}
For small $n$, equations~\ref{nz} and \ref{ntot} break down unless $f(z)$
describes a normal distribution.

\section{The form of $f(z)$}

Several considerations
suggest that $f(z)$ should be given by a power-law.  As a rough
estimate, we consider the volume affected by a star-forming
region to be that of the superbubble generated by its $N_*$ core-collapse
supernovae (SNe).  Following Oey \& Clarke (1997; hereafter OC97), the
mechanical power $L=N_* E_{\rm SN}/t_{\rm e}$, where $t_{\rm e}=40$
Myr is the life expectancy of $L$ for OB associations, and
$E_{\rm SN}=10^{51}$ erg is the individual SN energy.  The mechanical
luminosity function of $L$ is assumed to be a power-law with index $-\beta$
(OC97, eq.~1).  If the superbubbles grow until
they are confined by the ambient pressure, the standard, adiabatic
shell evolution implies that their final radii will be
related to $L$ as $R\propto L^{1/2}$ (OC97, eq.~29).
The total affected volume is therefore,
\begin{equation}
V_s\propto R^3 \propto L^{3/2} \quad .  
\end{equation}
In contrast, the total
mass of metals produced by $N_*$ SNe is given by,
\begin{equation}
M_z = m_y\ (t_{\rm e}/E_{\rm SN})\ L \quad ,
\end{equation}
where $m_y$ is the mean yield of metals per SN.  

Taking $z = M_z/V_s$, we find that $z\propto L^{-1/2}$ for
the superbubble population,
showing that the largest
superbubbles generate the lowest metallicities, owing to dilution into
larger volume.  The MDF for superbubbles is,
\begin{equation}
N_s(z)\ dz \propto N_s(R)\ dR\ \frac{dz}{dR} \propto z^{-3+2\beta} dz 
	\propto z\ dz \quad ,
\end{equation}
where $N_s(R)\propto R^{1-2\beta}$ is the size distribution for
objects at their final radii (OC97, eq.~41).  The final relation is
obtained for $\beta=2$, a value typical, perhaps, for all galaxies
(Oey \& Clarke 1998).  Here we see that there 
are larger numbers of metal-rich objects owing to the larger numbers
of smaller superbubbles.  Finally, the probability density for $z$ 
at any given point in the ISM is:
\begin{equation}
f(z) \propto \frac{4}{3}\pi R^3\ N_s(z) \propto	z^{-6+2\beta} \quad .
\end{equation}
For $\beta=2$, we obtain,
\begin{equation}\label{fz_pwr}
f(z)\ dz=C z^{-2}\ dz \ \ ,\quad z_{\rm min} < z < z_{\rm max} \quad .
\end{equation}
Since $f(z)$ is a probability density function, its integral must be unity,
and $C$ gives the appropriate normalization.
Thus, for a given generation of star formation, $f(z)$ is weighted
toward low-metallicity regions because of their large sizes.  The
limiting values $z_{\rm min}$ and $z_{\rm max}$ are determined by 
the corresponding $R_{\rm max}$ and $R_{\rm min}$.

For a Salpeter (1955)
IMF of slope --2.35 and SN progenitor masses between 8 and 120
$\msol$, we use $m_y=10\ \msol$, based roughly on models
by Woosley \& Weaver (1995).
We take $R_{\rm max} = 1300$ pc, a rather arbitrary value
corresponding to the characteristic parameter $R_{\rm e}$
of OC97; and $R_{\rm 
min} = 25$ pc, corresponding to individual SN remnants.
These yield $\log z_{\rm min}=-4.7$ and $\log z_{\rm max}=-3.0$, or,
taking O as a tracer of primary elements, minimum and maximum [O/H] of
--3.0 and --1.3 relative to solar.  The adopted yield of
$m_y=10\msol$ may be 
somewhat high (e.g., Woosley \& Weaver 1995), but the mean $f(z)$ may be
somewhat modified depending on $z_{\rm min}$ and $z_{\rm max}$.

\section{Results}

\begin{figure*}
\epsscale{1.8}
\plotone{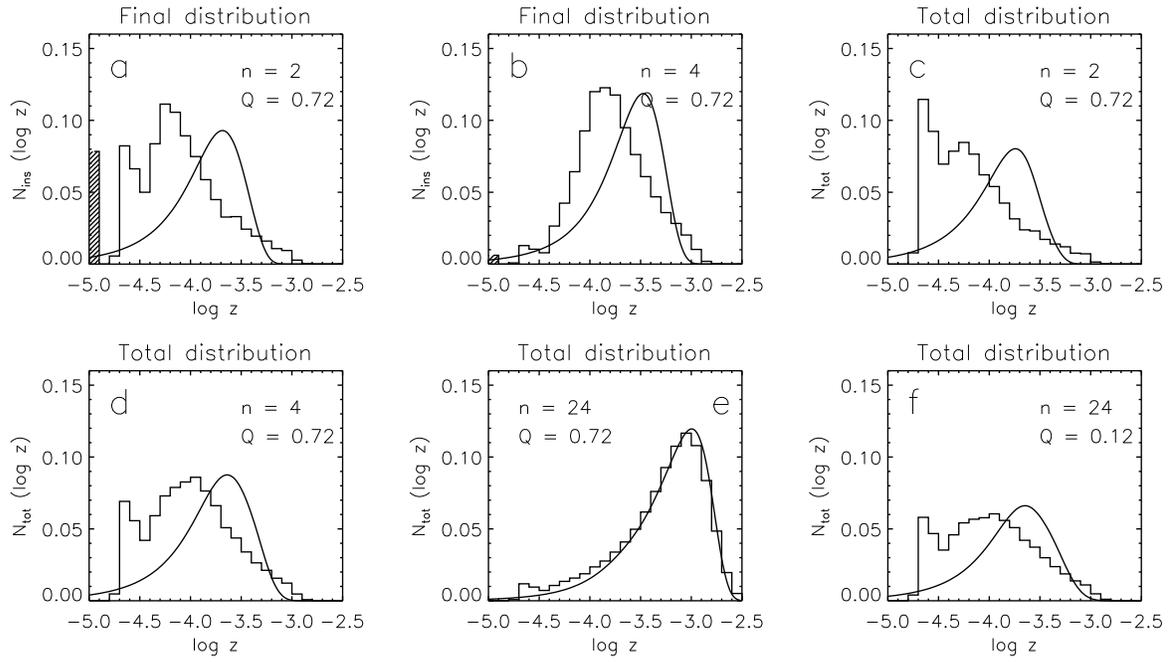}
\caption{Histograms show Monte Carlo models for MDFs,
and analytic curves show results from
equations~\ref{nz} and \ref{ntot}, with $n$ and $Q$ 
indicated.  The hatched bars in panels $a$ and $b$ show the proportion of
metal-free gas, placed arbitrarily in $\log z$.
\label{montecarlo}}
\end{figure*}

We can now construct Monte Carlo models of the MDF for small
$n$.  We generate the 
components $N_j(z)$ by drawing from $f(z)$ (equation~\ref{fz_pwr}) $j$
times and summing the drawn $z_i$ (equation~\ref{zsum}).  This is
repeated 5000 times to 
obtain distributions for $N_j(z)$.  Equations~\ref{nz_gen} and
\ref{ntot_gen} then provide the instantaneous and total MDFs.
The histograms in Figures~\ref{montecarlo}$a$ and $c$ show these
models of $\nins$ and $\ntot$ for $n=2$ and $Q=0.72$, and panels $b$
and $d$ show the same for $n=4$ generations.  The hatched bar in panels
$a$ and $b$ shows the fraction of primordial gas remaining, as determined by
$Q$.  The analytic curves show
the corresponding results from equations~\ref{nz} and \ref{ntot}.
After 2 generations at $Q=0.72$, we can still see the component
power-law distributions in the Monte Carlo models, and there is gross
deviation from the analytic approximation.  However, at only $n=4$,
we can already see how the stochastic model is approaching the
analytic version.  In panel $e$ at $n=24$, the models are in close agreement.

In contrast to this rapid evolution, Figure~\ref{montecarlo}$f$
demonstrates the effect of a small $Q$.  The model for $n=24,\ Q=0.12$
is similar to that in panel $d$ for $n=4,\ Q=0.72$.  Although these are
not identical, we can see that it is the product $nQ$ that
characterizes the evolutionary state, as a result of
equation~\ref{binomstats}.
Thus, {\it the relative filling factor of
contamination has the same importance as the number of contaminating
generations.}  While this statement may seem intuitively obvious, it
is worth emphasizing, since the implications are profound.

\begin{figure*}
\epsscale{1.8}
\plotone{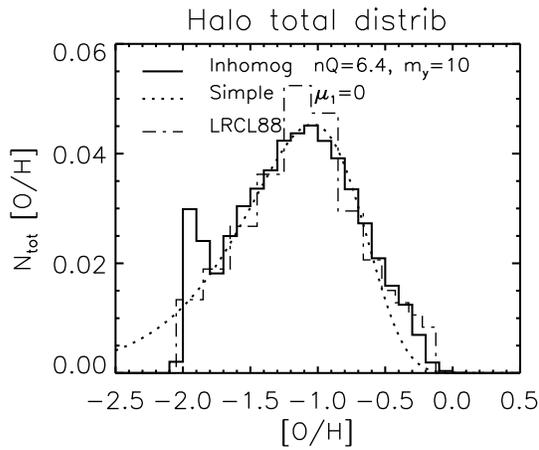}
\caption{MDF for Galactic halo stars from Laird {\etal}(1988; dash-dot line),
overplotted with a Monte Carlo inhomogeneous model for $n=8,\ Q=0.8$
(solid histogram).  The dotted line shows the Simple Model for 
$\log z_1/\zsol = -0.2$.  All models use $\mu_1 = 0.$
\label{halo}}
\end{figure*}

Figure~\ref{halo} shows data for the outer Galactic halo MDF 
(Laird {\etal}1988; dash-dot line), with
[Fe/H] converted to [O/H] following Pagel
(1989).  The solid histogram shows an inhomogeneous Monte Carlo
model for $nQ=6.4$.
There is remarkable similarity between the observations and our
rudimentary model, especially in the qualitative shape, having a
metal-rich tail and large ($\sim 1.0$ dex) dispersion.
The data also agree well, coincidentally,
with the Simple model (dotted line; e.g., Tinsley 
1980, equation~4.3) for present-day metallicity
$\log z_1/\zsol=-0.2$.  Both models take $\mu_1=0$.

\begin{figure*}
\epsscale{1.8}
\plotone{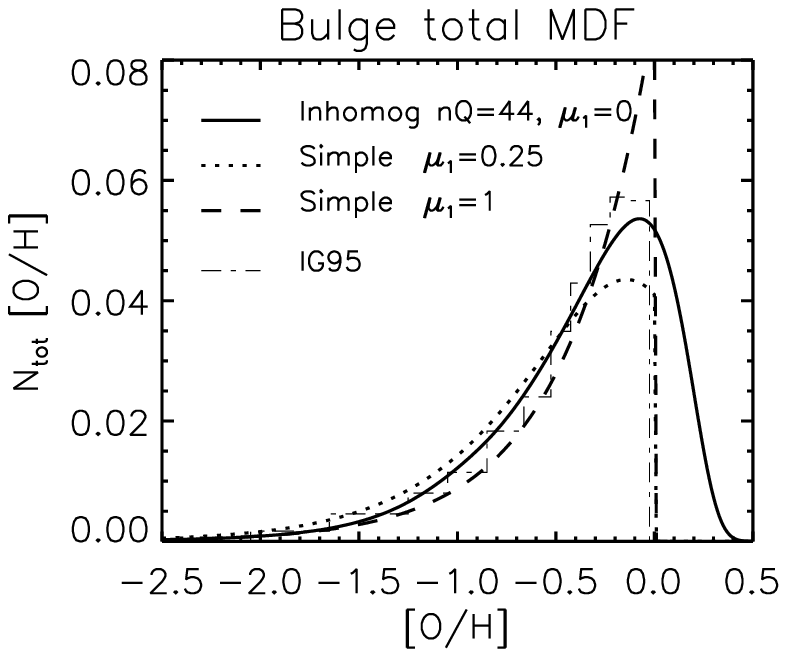}
\caption{Galactic bulge empirical F/G-dwarf MDF from Ibata \& Gilmore (1995,
dot-dashed line); they truncate data at [O/H]$>0$, due to uncertainties.
The solid line shows an analytic,
inhomogeneous model for $n=55,\ Q=0.8,\ \mu_1=0$ (solid
line).  Simple models are shown for $\mu_1=0.25,\ z_1=\zsol$ (dotted
line) and a non-consuming model, $\mu_1=1,\ z_1=\zsol$ (dashed line).
\label{bulge}}
\end{figure*}

However, it is essential to note that {\it the shapes of the two
models result from entirely different processes}.  The
high-metallicity turnover in the Simple model results purely from
consumption of gas into stars.  Figure~\ref{bulge} shows 
the Galactic bulge, a relatively evolved system, with data from Ibata
\& Gilmore (1995).  A Simple model having $z_1=\zsol$ requires 
$\mu_1\sim0.25$ to resemble the data (dotted line).  The dashed line
shows a similar homogeneous model with no gas consumption, i.e., 
a linear relation with equal numbers of stars created at all
metallicities ($\mu_1=1$).  The difference between these two
homogeneous models is thus entirely due to the depletion of gas into
stars, which also applies to the Simple model for the halo (Figure~\ref{halo}).

In contrast, the 
high-metallicity tail of the inhomogeneous halo model represents the
vestige of the $f(z)$ power-law; while the low-metallicity turnover
results from the Central Limit Theorem, as the component distributions
progress toward gaussians (cf. Figure~\ref{montecarlo}).  This offers an
alternative interpretation  
of the MDF vs. the Simple Model.  In addition, it reveals the
importance of the high-metallicity tail in relatively unevolved
systems, for probing the parent distribution $f(z)$.

An evolved inhomogeneous model like the Galactic bulge
should approach the same limit as the homogeneous
Simple model, since the metallicities contributed by $f(z)$
constitute a progressively smaller fraction of the
current $\bar{z}$ of the ISM (Edmunds 1975).  Figure~\ref{bulge}
indeed shows that all the models essentially coincide in the
low-metallicity tail.  However, at the high-metallicity end, the form
of the inhomogeneous model mainly reflects $\sigma_{\rm b}$ for the
most recent contaminating generations.  These become
progressively narrower (equation~\ref{binomstats}), but do not
truncate like the Simple model.

The inhomogeneous model matches the observed bulge MDF for
$\mu_1=0,\ z_1=\zsol$; while the Simple model would require $z_1$
to be 10 times larger for $\mu_1=0$, with almost half the distribution
at $0<$ [O/H] $\lesssim 1$.  Thus, it is especially
noteworthy that the inhomogeneous model {\it agrees with
both the Galactic halo and bulge MDFs by varying only the single
parameter, nQ.}  The predicted shape, dispersion, and mean $\bar{z}$
are linked, thus the simultaneous agreement 
for these features is highly encouraging.  It is also clear that the
supersolar MDF is a vital discriminant between the models.
Malinie {\etal}(1993) emphasize the 
importance of reproducing not only the low-metallicity tail, but also
the high-metallicity drop-off of the MDF.



\begin{figure*}
\epsscale{1.8}
\plotone{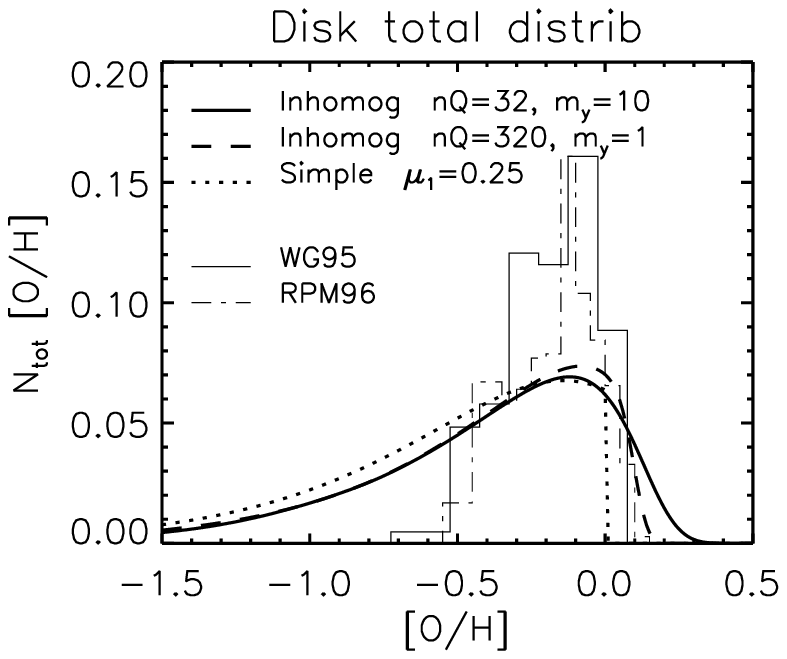}
\caption{MDF of Galactic disk F/G-dwarfs from Wyse \&
Gilmore (1995) and Rocha-Pinto \& Maciel (1996).  Analytic,
inhomogeneous models are shown for $n=40,\ Q=0.8,\ m_y=10\msol$ (solid
line) and $n=400,\ Q=0.8,\ m_y=1\msol$ (dashed line).  The dotted
curve shows the Simple Model for $z_1=\zsol$.  All models use $\mu_1=0.25$.
\label{gdwarf}}
\end{figure*}

Figure~\ref{gdwarf} shows histograms of the empirical
Galactic disk G-dwarf MDF for stars in solar neighborhood, with
[Fe/H] converted to [O/H] as before.  The solid curve
shows the analytic inhomogeneous model for $nQ=32,\
m_y=10\msol$; the dashed line shows a model with a lower yield
$m_y=1\msol$, which is thereby 10 times more evolved for the given
mean metallicity.  Finally, the dotted line shows the Simple model for
$z_1=\zsol$.  All models take $\mu_1 = 0.25$.  

The Simple model deviates most strongly from the data, illustrating
the ``G-dwarf Problem.''  Although the inhomogeneous models do not
solve the Problem, their MDFs are slightly more peaked, thus more
similar to the data.  Note also that Rocha-Pinto \& Maciel (1996)
truncated their derived distribution below an effective [O/H] $< -0.6$.
The predicted [O/H] FWHM dispersions in the present-day, instantaneous
MDF are 0.27 and 0.08 dex for the models with $nQ = 32$ and 320,
respectively.  These values can be compared to the present-day
observed dispersion in the solar neighborhood evidenced, for example,
by the well-known --0.3 dex deviation of the Orion nebula from solar
metallicity.

In summary, I present a rudimentary, one-zone model which considers
overlapping areas of contamination with no mixing.  This model thus
represents a limiting case opposite to the homogeneous Simple model.  An
under-appreciated point is that  the filling factor $Q$ of
contamination is as important as the number of star-forming
generations $n$.  The product $nQ$ therefore constitutes a 
single parameter that describes a system's evolutionary state.  $Q$
and $n$ may be independent, and are roughly associated with the global star
formation efficiency and age, respectively.  Thus a
system with low $Q$ can result in a present-day metal-poor ISM, but
having old stars (e.g., I~Zw~18); and a high $Q$ can yield an old, and
simultaneously, metal-rich population (e.g., the Galactic bulge).
By varying {\it only} the parameter $nQ$, the model can remarkably
match both the Galactic halo and bulge metallicity distributions, and
slightly improve the disk G-dwarf Problem.  As expected,
relatively unevolved systems are sensitive to
the parent $f(z)$.  For example, the high-metallicity tail of
the Galactic halo MDF is consistent with a power-law $f(z)$.
However, evolved systems are independent
of $f(z)$ and show progressively decreasing dispersions.
Further development of this inhomogeneous model, and investigation
of additional empirical constraints, are currently underway.

\acknowledgments

I am grateful to Cathie Clarke, Mike Fall, and Rosie Wyse for 
detailed discussions and suggestions, and also to Annette Ferguson
for early conversations.  I also enjoyed discussions with
many people during recent conferences. 


\end{document}